\documentclass[twocolumn]{aastex6}
\usepackage{natbib}
\usepackage{color}

\usepackage{color}
\usepackage{hyperref} %----set up hyper link to citations, equations...

\def \bea {\begin{eqnarray}}
\def \ena {\end{eqnarray}}                  
\def \bee {\begin{equation}}
\def \ene {\end{equation}}
\def    \simlt  {\lower.5ex\hbox{$\; \buildrel < \over \sim \;$}}
\def    \simgt  {\lower.5ex\hbox{$\; \buildrel > \over \sim \;$}}

\newcommand     \mum    {\,\mu{\rm m}}  % to use in math mode

\def	\cm		{\,{\rm {cm}}}
\def	\m		{\,{\rm m}}
\def	\km		{\,{\rm {km}}}

\def	\erg		{\,{\rm {erg}}}
\def	\eV		{\,{\rm {eV}}\,}
\def    \exp 		{\,{\rm {exp}}}
\def	\g		{\,{\rm g}}

\def	\K		{\,{\rm K}}

\def	\AU		{\,{\rm {AU}}}
\def	\pc		{\,{\rm {pc}}}

\def	\s		{\,{\rm s}}

\def    \yr  		{\,{\rm {yr}}}

\def	\AAt	 {\,{\rm \AA}}		% Angstrom
\def	\H		{\rm H}

\def    \apjl  		{\rm {ApJL}}

\def	\gas		{\rm {gas}}

\def    \abs     	{\rm {abs}}

\shorttitle{Destruction of H2-rich Objects}
\shortauthors{Hoang \& Loeb}

\begin{document}
\title{Destruction of molecular hydrogen ice and Implications for 1I/2017 U1 (`Oumuamua)}

\author{Thiem Hoang}
\affil{Korea Astronomy and Space Science Institute, Daejeon 34055, Republic of Korea; \href{mailto:thiemhoang@kasi.re.kr}{thiemhoang@kasi.re.kr}}
\affil{Korea University of Science and Technology, Daejeon 34113, Republic of Korea}
\author{Abraham Loeb}
\affil{Astronomy Department, Harvard University, 60 Garden Street, Cambridge, MA, USA; \href{mailto:aloeb@cfa.harvard.edu}{aloeb@cfa.harvard.edu}}

\begin{abstract}
The first interstellar object observed in our solar system, 1I/2017 U1 (`Oumuamua), exhibited a number of peculiar properties, including extreme elongation and acceleration excess. Recently, \cite{Seligman:2020vb} proposed that the object was made out of molecular hydrogen (H$_{2}$) ice. The question is whether H$_2$ objects could survive their travel from the  birth sites to the solar system. Here we study destruction processes of icy H$_2$ objects through their journey from giant molecular clouds (GMCs) to the interstellar medium (ISM) and the solar system, owing to interstellar radiation, gas and dust, and cosmic rays. We find that thermal sublimation due to heating by starlight can destroy `Oumuamua-size objects in less than 10 Myr. Thermal sublimation by collisional heating in GMCs could destroy H$_2$ objects of `Oumuamua-size before their escape into the ISM. Most importantly, the formation of icy grains rich in H$_2$ is unlikely to occur in dense environments because collisional heating raises the temperature of the icy grains, so that thermal sublimation rapidly destroys the H$_2$ mantle before grain growth.

\end{abstract}


\section{Introduction}\label{sec:intro}
The detection of the first interstellar object, 1I/2017 U1 (`Oumuamua) by the Pan-STARRS survey \citep{2017MPEC....U..181B} implies an abundant population of similar interstellar objects (\citealt{Meech:2017hu}; \citealt{2018ApJ...855L..10D}). An elongated shape of semi-axes $\sim 230\m\times 35\m$ is estimated from light-curve modeling (\citealt{2017ApJ...850L..36J}). The extreme axial ratio of $\gtrsim 5:1$ implied by `Oumuamua's lightcurve is mysterious (\citealt{Fraser:2018dg}; \citealt{2015MNRAS.454..593B}). 

\cite{2017ApJ...851L..38B} and \cite{Gaidos:2017wj} suggested that `Oumuamua is a contact binary, while others speculated that the bizarre shape might be the result of violent processes, such as collisions during planet formation. \cite{Domokos:2017tn} suggested that the elongated shape might arise from ablation induced by interstellar dust, and \cite{Hoang:2018es} suggested that it could originate from rotational disruption of the original body by mechanical torques. \cite{2019Icar..328...14S} suggested that the extreme elongation might arise from planetesimal collisions. The latest proposal involved tidal disruption of a larger parent object close to a dwarf star \citep{Zhang:2020eu}, but this mechanism is challenged by the preference for a disk-like shape implied by `Oumuamua's lightcurve (\citealt{2019MNRAS.489.3003M}).

Another peculiarity is the detection of non-gravitational acceleration in the trajectory of `Oumuamua \citep{Micheli:2018dl}. The authors suggested that cometary activity such as outgassing of volatiles could explain the acceleration excess. Interestingly, no cometary activity of carbon-based molecules was found by deep observations with the Spitzer space telescope \citep{2018AJ....156..261T} and Gemini North telescope \citep{Drahus:2018bd}. \cite{2018ApJ...868L...1B} explained the acceleration anomaly by means of radiation pressure acting on a thin lightsail, and \cite{MoroMartin:2019jf} and \cite{2019arXiv190500935S} suggested a porous object. \cite{Fitzsimmons:2017io} proposed that an icy object of unusual composition might survive its interstellar journey. Previously, \cite{2018A&A...613A..64F} suggested that `Oumuamua might be composed of H$_2$. However, \cite{Rafikov:2018} argued that the level of outgassing needed to produce the acceleration excess would rapidly change the rotation period of `Oumuamua, in conflict with the observational data.

Most recently, \cite{Seligman:2020vb} suggested hydrogen ice to explain `Oumuamua's excess acceleration and unusual shape. Their modeling implied that the object is $\sim$ 100 Myr old. Assuming a speed of $30\km\s^{-1}$, they suggested that the object was produced in a Giant Molecular Cloud (GMC) at a distance of $\sim 5$ kpc. However, their study did not consider the destruction of H$_2$ ice in the interstellar medium (ISM), but only through evaporation by sunlight. Here, we explore the evolution of H$_2$ ices from their potential GMC birth sites to the diffuse ISM and eventually the solar system. 

Assuming H$_2$ objects could be formed in GMCs by some mechanisms (\citealt{2016A&A...591A.100F}; \citealt{2018A&A...613A..64F}; \citealt{Seligman:2020vb}), we quantify their destruction and determine the minimum size of an H$_2$ object that can reach the solar system. We assume that the H$_2$ objects are ejected from GMCs into the ISM by some dynamical mechanism such as tidal disruption of bigger objects or collisions (see \citealt{2018ApJ...856L...7R}; \citealt{Rice:2019fc}). The evolution of H$_2$ objects in the ISM has additional implications for baryonic dark matter (\citealt{1996Ap&SS.240...75S}; \citealt{1999ApJ...516..195C}).

The structure of the paper is as follows. In Sections \ref{sec:destruct}-\ref{sec:ISM}, we calculate the destruction timescales from various processes for H$_2$ objects. In Section \ref{sec:results}, we compare the destruction times with the travel time for different object sizes. In Section \ref{sec:discuss}, we explore the formation of H$_2$-rich objects in dense GMCs and the implications for baryonic dark matter. We conclude with a summary of our main findings in Section \ref{sec:concl}. 

\section{Destruction of H$_2$ ice by interstellar radiation}\label{sec:destruct}
Let us first assume that H$_2$ objects could form in dense GMCs and get ejected into the ISM and examine several mechanisms for H$_2$ ice destruction. The binding energy of H$_2$ in the hydrogen ice is $E_{b}/k\sim 100\K$ (\citealt{1993ApJ...409L..65S}), equivalent to $E_{b}(\H_{2})\approx 0.01 \eV$. For simplicity, we assume a spherical object shape in our derivations, but the results can be easily generalized to other shapes. 
%{\bf The radius is assumed to be smaller than $R<R_{\rm grav}= (3E_{b}/(4\pi\rho_{\rm ice}m_{\H2}G)^{1/2}\approx 4143$ km, so that gravity is negligible compared to the binding energy of H$_2$.}

\subsection{Thermal sublimation}
Heating by starlight and Cosmic Microwave Background (CMB) radiation raises the surface temperature of H$_2$ ice. We assume that the local interstellar radiation field has the same spectrum as the interstellar radiation field (ISRF) in the solar neighborhood \citep{1983A&A...128..212M} with a total radiation energy density of $u_{\rm MMP}\approx 8.64\times 10^{-13} \erg\cm^{-3}$. We calibrate the strength of the local radiation field by the dimensionless parameter, $U$, so that the local energy density is $u_{\rm rad}=Uu_{\rm MMP}$. The CMB radiation is black body 
with a temperature $T_{\rm CMB}=2.725 (1+z)\K$ at a redshift $z$, and the radiation energy density is $u_{\rm CMB}=\int 4\pi B_{\nu}(T_{\rm CMB})d\nu/c=4\sigma T_{\rm CMB}^{4}/c\approx4.17\times10^{-13}(1+z)^{4}\erg\cm^{-3}$. At present, heating by the CMB is less important than by starlight.

The characteristic timescale for the evaporation of an H$_2$ molecule from a surface of temperature $T_{\rm ice}$ is
\bea
\tau_{\rm sub}=\nu_{0}^{-1}\exp\left(\frac{E_{b}}{k T_{\rm ice}}\right),\label{eq:tausub}
\ena
where $\nu_{0}$ is the characteristic oscillation frequency of the H$_2$ lattice \citep{1972ApJ...174..321W}. We adopt $\nu_0 = 10^{12} \s^{-1}$ for H$_2$ ice (\citealt{1986ApJ...303...56H}; \citealt{1993ApJ...409L..65S}).

Assuming that the H$_2$ ice has a layered structure, the sublimation rate for an H$_2$ object of radius $R$ is given by,
\bea
\frac{dR}{dt}=-\frac{\nu_{0}}{n_{\rm ice}^{1/3}}\exp\left(-\frac{E_{b}}{k T_{\rm ice}}\right),\label{eq:dasdt}
\ena
where $n_{\rm ice}\approx 3\times 10^{22}\cm^{-3}$ is the molecular number density of H$_2$ ice with a mass density of $\rho_{\rm ice}=0.1\g\cm^{-3}$. 

The sublimation time is then,
\bea
t_{\rm sub}(T_{\rm ice})=-\frac{R}{dR/dt}
= \frac{Rn_{\rm ice}^{1/3}}{\nu_{0}}\exp\left(\frac{E_{b}}{k T_{\rm ice}}\right),\label{eq:tsub}
\ena
where $dR/dt$ was substituted from Equation (\ref{eq:dasdt}).

Plugging the numerical parameters into the above equation, we obtain,
\bea
t_{\rm sub}(T_{\rm ice})\simeq 2.95\times 10^{7}\left(\frac{R}{1\km}\right)\exp\left(100\K\left[\frac{1}{T_{\rm ice}}-\frac{1}{3\K}\right]\right) \rm yr\label{eq:tsub1}~~~~
\ena
for H$_2$ ice, and the ice temperature is normalized to $T_{\rm ice}=3\K$ expecting that starlight raises the surface temperature above $T_{\rm CMB}$. At the minimum temperature of the present-day CMB radiation, $T_{\rm obj}= 2.725$ K, the sublimation time is $t_{\rm sub}\approx 0.85$ Gyr for $R= 1\km$. 

The heating rate due to absorption of isotropic interstellar radiation and CMB photons is given by,
\bea
\frac{dE_{\abs}}{dt}=\pi R^{2}c(Uu_{\rm MMP}+u_{\rm CMB})\epsilon_{\star},\label{eq:dErad}
\ena
where $\epsilon_{\star}$ is the surface emissivity averaged over the background radiation spectrum. 

The cooling rate by thermal emission is given by,
\bea
\frac{dE_{\rm emiss}}{dt}=4\pi R^{2}\epsilon_{T}\sigma T^{4},\label{eq:dErad}
\ena
where $\epsilon_{T}=\int d\nu \epsilon(\nu)B_{\nu}(T)/\int d\nu B_{\nu}(T)$ is the bolometric emissivity, integrated over all radiation frequencies, $\nu$.

The energy balance between radiative heating and cooling yields the surface equilibrium temperature,
\bea
T_{\rm ice}&=&\left(\frac{c(Uu_{\rm MMP}+u_{\rm CMB})}{4\sigma}\right)^{1/4}\left(\frac{\epsilon_{\star}}{\epsilon_{T}}\right)^{1/4}
\nonumber\\
&\simeq& 3.59\left(U+[1+z]^{3}\right)^{1/4}\left(\frac{\epsilon_{\star}}{\epsilon_{T}}\right)^{1/4}\K.~~~~
\ena
At this temperature, the sublimation time is short, less than $\sim 1 \times 10^{5}$ yr, according to Equation (\ref{eq:tsub1}). 

However, to access the actual temperature of the ice, we need to take account of evaporative cooling (\citealt{1972ApJ...174..321W}; \citealt{2015ApJ...806..255H}). The cooling rate by evaporation of H$_2$ is given by,
\bea
\frac{dE_{\rm evap}}{dt}=\frac{E_{b}dN_{\rm mol}}{dt}=\frac{E_{b}N_{s}}{\tau_{\rm sub}(T_{\rm ice})},
\ena
where $dN_{\rm mol}/dt$ is the evaporation rate, namely, the number of molecules evaporating per unit time, and $N_{s}=4\pi R^{2}/r_{s}^{2}$
is the number of surface sites with $r_{s}=10~{\rm \AA}$ being the average size of the H$_2$ surface site (\citealt{1993ApJ...409L..65S}). 

The ratio of evaporative to radiative cooling rates is given by,
\bea
\frac{dE_{\rm evap}/dt}{dE_{\rm emiss}/dt}&=&\frac{E_{b}\nu_{0}\exp(-E_{b}/kT_{\rm ice})}{\epsilon_{T}\sigma T_{\rm ice}^{4}r_{s}^{2}}\nonumber\\
&\simeq& \left(\frac{1.1}{\epsilon_{T}}\right)\left(\frac{3\K}{T_{\rm ice}}\right)^{4}\exp\left(100\K\left[\frac{1}{T_{\rm ice}}-\frac{1}{3\K}\right]\right)\nonumber\\
&\times&\left(\frac{E_{b}}{0.01\eV}\right)\left(\frac{10\AAt}{r_{s}}\right)^{2},\label{eq:evap_emiss}
\ena
implying that the evaporative cooling dominates over radiative cooling for $T_{\rm ice}\gtrsim 3\K$. %Therefore, the H$_2$ surface temperature is maintained at $T_{\rm ice}\approx 3.06\K$.

For an H$_2$ object moving at a speed, $v_{\rm obj}$, through the ISM, the heating rate by gas collisions is given by,
\bea
\frac{dE_{\rm coll}}{dt}=\frac{1}{2}\pi R^{2} n_{\H}\mu m_{\H}v_{\rm obj}^{3},
\ena
where $\mu$ is the mean molecular weight of the ISM and $m_{\H}$ is the mass of a hydrogen atom. For the cosmic He abundance, $\mu=1.4$.

The ratio of collisional heating to radiative heating by starlight is given by, 
\bea
%\frac{dE_{\rm coll}}{dE_{\rm abs}}&=&\frac{(n_{\H}\mu m_{\H}v_{\rm obj}^{3}/2)}{4cUu_{\rm MMP}\epsilon_{\star}}\nonumber\\
%&\simeq& 3\left(\frac{n_{\H}}{10^{4}\cm^{-3}}\right)\left(\frac{v_{\rm obj}}{30\km\s^{-1}}\right)^{3}\left(\frac{U}{\epsilon_{\star}}\right),
\frac{dE_{\rm coll}}{dE_{\rm abs}}&=&\frac{(n_{\H}\mu m_{\H}v_{\rm obj}^{3}/2)}{cUu_{\rm MMP}\epsilon_{\star}}\nonumber\\
&\simeq& 1.2\left(\frac{n_{\H}}{10^{3}\cm^{-3}}\right)\left(\frac{v_{\rm obj}}{30\km\s^{-1}}\right)^{3}\left(\frac{U}{\epsilon_{\star}}\right),
\ena
implying dominance of collisional heating if $n_{\H}\gtrsim cUu_{\rm MMP}/(\mu m_{\H}v_{\rm obj}^{3}/2)\simeq 825U(30\km\s^{-1}/v_{\rm obj})^{3}\cm^{-3}$, assuming $\epsilon_{\star}=1$. Thus, in GMCs, collisional heating is important and can destroy H$_2$ objects rapidly (see Section \ref{sec:results}). For the diffuse ISM, collisional heating is negligible.

The final energy balance equation reads
\bea
\frac{dE_{\rm abs}}{dt}+\frac{dE_{\rm coll}}{dt}=\frac{dE_{\rm emiss}}{dt}+\frac{dE_{\rm evap}}{dt}.\label{eq:balance}
\ena

We numerically solve the above equation for the equilibrium temperature, and obtain $T_{\rm ice}=2.996\approx 3\K$, assuming $U=1$, the present CMB, and $n_{\rm H}=10\cm^{-3}$. At this temperature, Equation (\ref{eq:tsub1}) implies the sublimation time of $\sim 30$ Myr for $R=1$ km. Passing near a region with enhanced radiation fields, e.g., near a star, would reduce the sublimation time significantly.

\subsection{Photodesorption}
Next we estimate the lifetime of an icy H$_2$ object to UV photodesorption. Let $Y_{\rm pd}$ be the photodesorption yield, defined as the number of molecules ejected over the total number of incident UV photons. The rate of mass loss due to UV photodesorption is
\bea
\frac{dm}{dt} = \frac{4\pi R^{2} \rho_{\rm ice} dR}{dt}=-\bar{m}Y_{\rm pd}F_{\rm UV}\pi a^{2},
\ena
where $\bar{m}$ is the mean mass of ejected molecules, and $F_{\rm UV}$ is the flux of UV photons. This yields
\bea
\frac{dR}{dt} &=& -\frac{\bar{m}Y_{\rm pd}F_{\rm UV}}{4\rho_{\rm ice}}\nonumber\\
&\simeq& -262\left(\frac{F_{\rm UV}}{10^{7}\cm^{-2}\s^{-1}}\right)\left(\frac{Y_{\rm pd}}{10^{3}}\right)\left(\frac{0.1\g\cm^{-3}}{\rho_{\rm ice}}\right) \frac{\AAt}{\yr},~~~~~
\ena
where $\bar{m}=2m_{\H}$, $Y_{\rm pd}=h\nu/E_{b}=10^{3}$ for $h\nu= 10 \eV$, and $F_{\rm UV}= 10^{7}\cm^{-2}\s^{-1}$ for the ISRF.

We define $G=F_{\rm UV}/F_{\rm UV,MMP}$ to calibrate the strength of background UV radiation, where $F_{\rm UV,MMP}=10^{7}\cm^{-2}\s^{-1}$ is the UV flux of the standard ISRF. The photodesorption time for an object of radius $R$ is,
\bea
t_{\rm pd}&=&-\frac{R}{dR/dt}\nonumber\\
&\simeq& \frac{7.5\times 10^{10}}{G}\left(\frac{R}{1\km}\right) \left(\frac{10^{3}}{Y_{\rm pd}}\right)\left(\frac{10^{7}\cm^{-2}\s^{-1}}{F_{\rm UV,MMP}}\right)\rm yr.~~~~~
\ena
An enhancement of the local UV radiation near an OB association can increase the photodesorption rate by a factor of $G$.

\section{Destruction by Cosmic Rays}\label{sec:CRs}
The stopping power of a relativistic proton in H$_2$ ice is, $dE/dx\sim -10^{6} \eV\cm^{-1}$ at an energy $E\sim 1$ GeV (\citealt{2015ApJ...806..255H}; \citealt{2017ApJ...837....5H}). The corresponding penetration length is $R_{p}=-E/(dE/dx)\sim 10^{3} \cm=10$ m.

The ice volume evaporated by a cosmic ray (CR) proton is determined by the heat transfer from the CR to the ice volume that reaches an evaporation temperature $T_{\rm evap}\sim E_{b}/3k$ (i.e., thermal energy per H$_2$ comparable to the binding energy). Since the object radius is much larger than the above penetration length, the volume of ice evaporated by a CR proton, $\delta V$, is given by,
\bea
n_{\rm ice}\delta V E_{b}=E_{\rm CR}.
\ena

Because the penetration length is much shorter than the `Oumuamua's estimated size, CRs would gradually erode the object. The fraction of the object volume eroded by CRs per unit of time is.
\bea
\frac{1}{V}\frac{dV}{dt}=-\frac{4\pi R^{2}F_{\rm CR}\delta V}{V}=-\frac{3F_{\rm CR}E_{\rm CR}}{Rn_{\rm ice}E_{b}},
\ena
where $V=4\pi R^{3}/3$ is the object's volume.

The timescale required to eliminate the object is,
\bea
t_{\rm CR}&=&-\frac{V}{dV/dt}=\frac{Rn_{\rm ice}E_{b}}{3F_{\rm CR}E_{\rm CR}}\nonumber\\
&\simeq& 3.2\times 10^{8}\left(\frac{R}{1\km}\right)\left(\frac{E_{b}}{0.01\eV}\right)\left(\frac{10^{9}\eV}{E_{\rm CR}}\right)\nonumber\\
&\times&\left(\frac{1\cm^{-2}\s^{-1}}{F_{\rm CR}}\right)\yr,~~~~~
\ena
where $F_{\rm CR}=1\cm^{-2}\s^{-1}$ is the flux of proton CRs of $E=1$ GeV. The above result is comparable to the estimate by \cite{1996Ap&SS.240...75S}.

%Assuming the flux $F_{\rm CR}\sim 1 proton \cm^{-2}\s^{-1}$, \cite{1996Ap&SS.240...75S} derived the sublimation timescale of an icy object of radius $L$ by CRs sublimation as
%\bea
%t_{\rm CR}=7.4\times 10^{7}
%\left(\frac{R}{100m}\right) \rm yr.
%\ena

%Following \cite{1985A&A...144..147L}, the flux of CRs in the ISM is given by
%\bea
%F_{\rm CR}\sim 0.3 proton \cm^{-2}\s^{-1}.
%\ena
%The mean free path of protons in H$_2$ ice is $\lambda_{mfp}\sim 6m$.

The contribution of heavy ion CRs is less important than proton CRs because their flux is lower; for iron ions, the abundance ratio is $F_{\rm Fe}/F_{p}= 1.63\times 10^{-4}$ (see \citealt{1985A&A...144..147L}).

\section{Destruction by interstellar matter}\label{sec:ISM}
\subsection{Nonthermal Sputtering}
At a characteristic speed of $v_{\rm obj}\sim 30 \km\s^{-1}$, each ISM proton delivers an energy of $E_{p}=m_{\H}v_{\rm obj}^{2}/2 \approx 4.66$ eV to the impact location. Thus, protons can eject H$_2$ out of the ice surface with a sputtering yield of $Y_{\rm sp}\sim E_{p}/E_{b}\sim 460$.

The destruction time of H$_2$ ice by sputtering is given by,
\bea
t_{\rm sp}&=&-\frac{R}{dR/dt}=\frac{4\rho_{\rm ice} R}{n_{\H}m_{\H}v_{\rm obj}Y_{\rm sp}}\nonumber\\
&\simeq& 2.6\times 10^{10}\left(\frac{0.1\g\cm^{-3}}{\rho_{\rm ice}}\right)\left(\frac{R}{1\km}\right)\left(\frac{10\cm^{-3}}{n_{\H}}\right)\nonumber\\
&\times& \left(\frac{30\km\s^{-1}}{v_{\rm obj}}\right)
 \left(\frac{10^{3}}{Y_{\rm sp}}\right) \rm yr,\label{eq:tau_sp}
\ena
which is short only in GMCs and unimportant for the diffuse ISM. Moreover, most of proton's energy may go into forming a deep track instead of surface heating, reducing the sputtering effect.

\subsection{Impulsive collisional heating and transient evaporation}
Collisions of H$_2$ ice with the ambient gas at high speeds can heat the frontal area to a temperature $T_{\rm evap}$, resulting in transient evaporation. The volume of ice evaporated by a single collision, $\delta V$, can be given by
\bea
n_{\rm ice} \delta V E_{b}=\frac{1}{2}\mu m_{\H}v_{\rm obj}^{2}, 
\ena
where the impact kinetic energy is assumed to to be fully converted into heating.

The evaporation rate by gas collisions is given by,
\bea
\frac{1}{V}\frac{dV}{dt}=-\frac{\pi R^{2}n_{\H}v_{\rm obj}\delta V}{V}=-\frac{3n_{\H}\mu m_{\H}v_{\rm obj}^{3}}{8Rn_{\rm ice}E_{b}}.
\ena

The evaporation time by gas collisions is then,
\bea
t_{\rm evap,gas}&=&-\frac{V}{dV/dt}=\frac{8Rn_{\rm ice}E_{b}}{3n_{\H}\mu m_{\H}v^{3}}\nonumber\\
&\simeq& 6.5\times 10^{9}\left(\frac{R}{1\km}\right)\left(\frac{30\km\s^{-1}}{v_{\rm obj}}\right)^{3}\nonumber\\
&\times&\left(\frac{n_{\H}}{10\cm^{-3}}\right)^{-1}\left(\frac{E_{b}}{0.01\eV}\right)\yr,\label{eq:tevap_gas}
\ena
somewhat shorter than the sputtering time given in Equation (\ref{eq:tau_sp}).

Similarly, dust grains of mass $m_{gr}$ deposit a kinetic energy of $E_{\rm gr}=m_{\rm gr}v_{\rm obj}^{2}/2$ upon impact, resulting in transient evaporation. The evaporation rate by dust collisions is given by
\bea
\frac{1}{V}\frac{dV}{dt}=-\frac{\pi R^{2}n_{\rm gr}v_{\rm obj}\delta V}{V}=-\frac{3n_{\rm gr}m_{\rm gr}v_{\rm obj}^{3}}{8Rn_{\rm ice}E_{b}},
\ena
yielding a dust evaporation time,
\bea
t_{\rm evap,d}=-\frac{V}{dV/dt}=\frac{8Rn_{\rm ice}E_{b}}{3n_{\rm gr}m_{\rm gr}v_{\rm obj}^{3}}.\label{eq:tevap_dust}
\ena

Assuming that all grains have the same size, $a$, and using the dust-to-gas mass ratio $M_{d/g}= n_{\rm gr}4\pi a^{3}\rho_{\rm gr}/(3\mu m_{\H}n_{\H})$, one obtains the grain number density,
\bea
n_{\rm gr}&=&\frac{M_{d/g}(3\mu m_{\H}n_{\H})}{4\pi a^{3}\rho_{\rm gr}}\nonumber\\
&\approx& 1.85\times 10^{-11}\left(\frac{M_{d/g}}{100}\right)\left(\frac{n_{\H}}{10\cm^{-3}}\right)\nonumber\\
&\times&\left(\frac{0.1\mum}{a}\right)^{3}\cm^{-3},\label{eq:ngr}
\ena
where $\rho_{\rm gr}=3\g\cm^{-3}$ is assumed.

Substituting $n_{\rm gr}$ into Equation (\ref{eq:tevap_dust}) yields,
\bea
t_{\rm evap,d}&\simeq& 6.5\times 10^{11}\left(\frac{R}{1\km}\right)\left(\frac{30\km\s^{-1}}{v_{\rm obj}}\right)^{3}\nonumber\\
&\times&\left(\frac{n_{\H}}{10\cm^{-3}}\right)\left(\frac{E_{b}}{0.01\eV}\right)\yr.~~
\ena

The destruction by dust is less efficient than by gas due to a lower dust mass. 
%The situation is different for the water ice because gas collisions is not efficient to sublimate water ice at speeds $v=30km/s$.

\subsection{Destruction by bow shocks}
For a cold GMC of temperature $T_{\gas}\sim 3\K$, the thermal velocity of the gas is $v_{\rm T}=(3kT_{\gas}/m_{\H})^{1/2}\sim 0.27 (T_{\gas}/3\K)^{1/2} \km\s^{-1}$. Objects moving rapidly through the gas with $v_{\rm obj}\sim 30\km\s^{-1}\gg v_{T}$, will produce a bow shock if their radius is larger than the mean free path of gas molecules (\citealt{Landau:1959vb}). The mean free path for molecular collisions is $\lambda_{\rm mfp}\sim 1/(n_{\H}\sigma_{\H2})= 10^{4}(10^{6}\cm^{-3}/n_{\H}) (10^{-15}\cm^{-2}/\sigma_{\H2}) \km$ with $\sigma_{\H2}$ being the H$_2$ cross-section. Thus, for objects larger than $R= 10^{4}$ km, bow shocks are formed if the gas density $n_{\H}\gtrsim 10^{6}\cm^{-3}$. The post-shock gas has a high temperature and can be efficient in thermal sputtering. However, bow shocks are not expected to form for objects of $R<10^{4}\km$ and $n_{\H}<10^{6}\cm^{-3}$.
%R> mfp because it is the minimum scale below which gas species share energy via collisions and behave like a fluid 

\section{Numerical Results}\label{sec:results}

\subsection{Destruction in the ISM}
Assuming that H$_2$ objects of various sizes are produced in a nearby GMC, we estimate the minimum size of objects that could reach the Earth. The closest GMC, W51, is located at a distance of $D_{\rm GMC}=5.2$ kpc. Thus, at a speed of $30 \km\s^{-1}$, it takes $t_{\rm trav}\approx 1.6\times 10^{8}$ yr for objects to reach the solar system.

\begin{figure}
\includegraphics[width=0.5\textwidth]{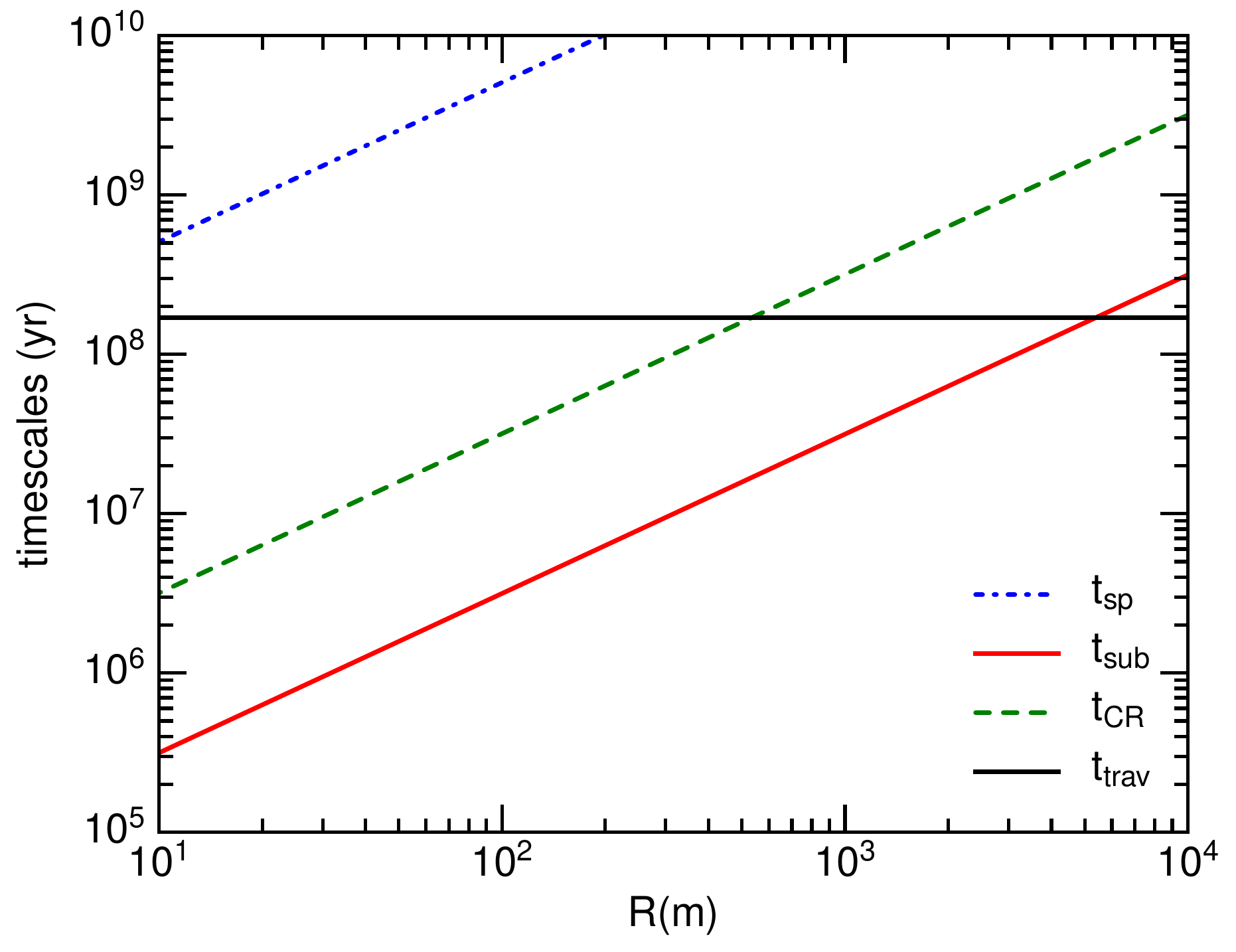}
\caption{Comparison of various destruction timescales (slanted colored lines) as a function of the object radius (in meters) to the travel time from a GMC at a distance of 5.2 kpc, assuming a characteristic speed of 30 $\km\s^{-1}$ (horizontal black line).}
\label{fig:tcom}
\end{figure}

Figure \ref{fig:tcom} compares the various destruction times with $t_{\rm trav}$ for different object radii at a typical speed. The sublimation time is obtained using the equilibrium temperature $T_{\rm ice}=2.9939\approx 3\K$ (see Section \ref{sec:destruct}). We find that only very large objects of radius $R>5\km$ could survive thermal sublimation and reach the solar system. The minimum size of the objects that can survive is obtained by setting $t_{\rm sub}=t_{\rm trav, ISM}$, yielding
\bea
R_{\rm min, ISM}&=&5.4\left(\frac{D_{\rm GMC}}{5.2~\rm kpc}\right)\left(\frac{30\km\s^{-1}}{v_{\rm obj}}\right)\times
\nonumber\\
&&\exp\left(100\K\left[\frac{1}{T_{\rm ice}}-\frac{1}{3\K}\right]\right)\km.\label{eq:Rmin}
\ena

\subsection{Destruction on the way from the center of GMCs to the ISM}
The total gas column density toward the densest GMC amounts to extinction of $A_{V}\sim 200$ (see e.g., \citealt{1983A&A...128..212M}), which corresponds to a hydrogen column density of $N_{\H}\sim 3.74\times 10^{23}\cm^{-2}$ based on the scaling $N_{\H}/A_{V}\approx (5.8\times 10^{21}/R_{V})\cm^{-2}/{\rm mag}$ with $R_{V}=3.1$ (\citealt{2011piim.book.....D}). Assuming a mean GMC density $n_{\H}=10^{4}\cm^{-3}$ and a GMC radius $R_{\rm GMC}\approx 12\pc$, the travel time is $t_{\rm trav, GMC}=R_{\rm GMC}/v_{\rm obj}\approx 3.91\times 10^{5}(R_{\rm GMC}/12\pc)(30\km\s^{-1}/v_{\rm obj})\yr$.

Equation (\ref{eq:tevap_gas}) implies a destruction time by gas collisions of $t_{\rm evap,gas}\sim 9\times 10^{6}(R/1\km)(n_{\H}/10^{4}\cm^{-3})\yr$, which is longer than the travel time.

As shown in Section \ref{sec:destruct}, collisional heating is important in GMCs because of their high density, $n_{\H}\gtrsim 10^{3}\cm^{-3}$. Assuming that a fraction $\eta$ of the impinging proton's energy is converted into surface heating to a temperature below $T_{\rm evap}$, collisional heating raises the temperature of the surface to:
\bea
T_{\rm ice}&=&\left(\frac{\eta n_{\H}1.4m_{\H}v_{\rm obj}^{3}/2}{4\sigma \epsilon_{T}}\right)^{1/4}\nonumber\\
&\simeq& 6.1\left(\frac{n_{\H}}{10^{4}\cm^{-3}}\right)^{1/4}\left(\frac{\eta}{\epsilon_{T}}\right)^{1/4}\left(\frac{v_{\rm obj}}{30\km\s^{-1}}\right)^{3/4}\K,~~~~~\label{eq:Tgr}
\ena
which implies effective cooling by evaporation (see Equation \ref{eq:evap_emiss}). 

To find the actual equilibrium temperature, we solve Equation (\ref{eq:balance}) and obtain $T_{\rm ice}\approx 3.26\K$ for $n_{\H}=10^{4}\cm^{-3}$, assuming $\eta=\epsilon_{T}=1$. Substituting this typical temperature into Equation (\ref{eq:tsub}) yields,
\bea
t_{\rm sub}(T_{\rm ice})&\simeq& 2.1\times 10^{6}\left(\frac{R}{1\km}\right)\nonumber\\
&&\times\exp\left(100\K\left[\frac{1}{T_{\rm ice}}-\frac{1}{3.26\K}\right]\right) \yr.\label{eq:tsub_GMC}
\ena

The minimum size of the objects that can survive is obtained by setting $t_{\rm sub}=t_{\rm trav, GMC}$, yielding
\bea
R_{\rm min, GMC}&\approx &186\left(\frac{R_{\rm GMC}}{12\pc}\right)\left(\frac{30\km\s^{-1}}{v_{\rm obj}}\right)\times
\nonumber\\
&&\exp\left(100\K\left[\frac{1}{T_{\rm ice}}-\frac{1}{3.26\K}\right]\right)~\rm m.\label{eq:Rmin}
\ena

We conclude that H$_2$ objects cannot survive their journey from their GMC birthplace to the ISM if their radius is below $R_{\rm min, GMC}$. The actual value $R_{\rm min, GMC}$ would be larger because the surface temperature is higher when objects are moving through the core of GMCs with higher gas density.

\subsection{Destruction in the solar system}
When `Oumuamua entered the solar system, solar radiation heated the frontal surface to an equilibrium. We numerically solve Equation (\ref{eq:balance}) and obtain the equilibrium temperatures $T_{\rm ice}= 7.94$ and $7.15\K$ at heliocentric distances $D=0.25$ and $0.5\AU$. Substituting these temperatures into Equation (\ref{eq:tsub1}), one obtains $t_{\rm sub}= 3.18$ and $12.72$ days,
assuming $R=300\m$. These sublimation times are shorter than the travel time of $t_{\rm trav}=0.5\AU/v_{\rm obj}=14.42$ days. The sputtering by the solar wind is less important than thermal sublimation.

\section{Discussion}\label{sec:discuss}

\subsection{Can H$_2$-rich grains form in dense clouds?}
\cite{Seligman:2020vb} suggested that H$_2$ ice objects can form by means of accretion and coagulation of dust grains in the densest region of a GMC where the gas density is $n_{\H}\sim 10^{5}\cm^{-3}$ and temperature is $T_{\gas}\sim 3\K$. Below, we show that H$_2$-rich grains cannot form in the GMC due to destruction by collisional heating, preventing the formation of H$_2$ objects. 

At low temperatures, the accretion of H$_2$ molecules from the gas phase onto a grain core is a main process enabling the formation of an H$_2$ mantle. The characteristic timescale for forming an icy grain of radius $a$ is given by,
\bea
t_{\rm acc}&=&\frac{m_{\rm gr}}{1.05s_{\H}n_{\H}m_{\H}v_{\rm th}\pi a^{2}}=\frac{4\rho_{\rm ice} a}{3.15s_{\H}n_{\H}m_{\H}v_{\rm th}}\nonumber\\
&\simeq& 10^{2}\left(\frac{a}{1\mum}\right)\left(\frac{n_{\H}}{10^{5}\cm^{-3}}\right)^{-1}\left(\frac{T_{\gas}}{T_{\rm CMB}}\right)^{1/2}\nonumber\\
&\times&\left(\frac{\rho_{\rm ice}}{0.1\g\cm^{-3}}\right)\yr,\label{eq:tacc}~~~
\ena
where the thermal velocity $v_{\rm th}=(8kT_{\gas}/\pi m_{\H})^{1/2}$, the factor $1.05$ accounts for $n(\rm He)/n_{\H}=0.1$, and the sticking coefficient $s_{\H}=1$ is assumed.
% Here, vth defined as the mean of the magnitude of the velocity, like the speed of light

The timescale to form H$_2$ ice from collisions between two icy grains of equal sizes $a$ and relative velocity $v_{gg}$ is given by,
\bea
t_{\rm coag}&=&\frac{1}{n_{\rm gr}v_{gg}\pi a^{2}}=\frac{4a\rho_{\rm gr}}{3M_{d/g}\mu m_{\H} n_{\H}v_{gg}s_{\rm gr}}\nonumber\\
&\simeq& \frac{2.5\times 10^{5}}{s_{\rm gr}}\left(\frac{a}{1\mum}\right)\left(\frac{10^{5}\cm^{-3}}{n_{\H}}\right)\nonumber\\
&\times &\left(\frac{0.1\km\s^{-1}}{v_{gg}}\right)\left(\frac{0.01}{M_{d/g}}\right) \yr,
\ena
amounting to $\sim 10^{4}$ yr for a density of $n_{\H}\sim 10^{6}\cm^{-3}$, a sticking coefficient, $s_{\rm gr}= 1$, and the grain number density, $n_{\rm gr}$, is given by Equation (\ref{eq:ngr}). In conclusion, the timescale to form micron-sized grains by coagulation is much longer than the formation time by gas accretion, in agreement with the estimate by \cite{Seligman:2020vb}. 

However, \cite{Seligman:2020vb} did not consider the destructive effect of icy H$_2$ grains by collisional heating by gas. \cite{1969Natur.224..251G} noted that, at a density of $n_{\H}>10^{5}\cm^{-3}$, collisional heating might prevent the formation of H$_2$ ice. We calculate the grain temperature heated by gas with a minimum temperature $T_{\gas}=T_{\rm CMB}=2.725\K$ as follows:
\bea
%T_{\rm gr}&=&\left(\frac{n_{\H}1.4m_{\H} v_{\rm th}^{3}}{2\sigma \langle Q_{\rm abs}\rangle_{T}}\right)^{1/4}\nonumber\\
T_{\rm gr}&=&\left(\frac{1.05n_{\H}v_{\rm th}\times \pi a^{2}\times 2kT_{\rm gas}}{4\pi a^{2}\sigma \langle Q_{\rm abs}\rangle_{T}}\right)^{1/4}\nonumber\\
&\simeq& 3.02\left(\frac{n_{\H}}{10^{5}\cm^{-3}}\right)^{1/4}\left(\frac{T_{\gas}}{T_{\rm CMB}}\right)^{3/8}\left(\frac{10^{-4}}{\langle Q_{\rm abs}\rangle_{T}}\right)^{1/4}\K,~~~~~\label{eq:Tgr}
\ena
where $2kT_{\rm gas}$ is the mean kinetic energy of thermal particles colliding with the grain, and $\langle Q_{\rm abs}\rangle_{T}\approx 1.1\times 10^{-4}(a/1\mum)(T/3\K)^{2}$ for silicate grains (\citealt{2011piim.book.....D}). Gas collisions eventually lead to thermal equilibrium at $T_{\rm gr}=T_{\gas}$.

Substituting this typical temperature and the grain size $a=1\mum$ into Equation (\ref{eq:tsub}) yields,
\bea
%t_{\rm sub}(T_{\rm ice})\simeq 0.029\left(\frac{a}{1\mum}\right)\exp\left(100\K\left[\frac{1}{T_{\rm ice}}-\frac{1}{3\K}\right]\right) \yr,
t_{\rm sub}(T_{\rm gr})\simeq 0.85\left(\frac{a}{1\mum}\right)\exp\left(100\K\left[\frac{1}{T_{\rm gr}}-\frac{1}{T_{\rm CMB}}\right]\right) \yr,
\label{eq:tsub1_GMC}~~~~
\ena
much shorter than the accretion time $t_{\rm acc}$ given in Equation (\ref{eq:tacc}). The gas temperature in realistic GMCs is larger than $T_{\rm CMB}$ due to CR heating, resulting in a much shorter sublimation time. We therefore conclude that micron-sized H$_2$ grains cannot form in dense GMCs due to collisional heating.

\begin{figure}
\includegraphics[width=0.5\textwidth]{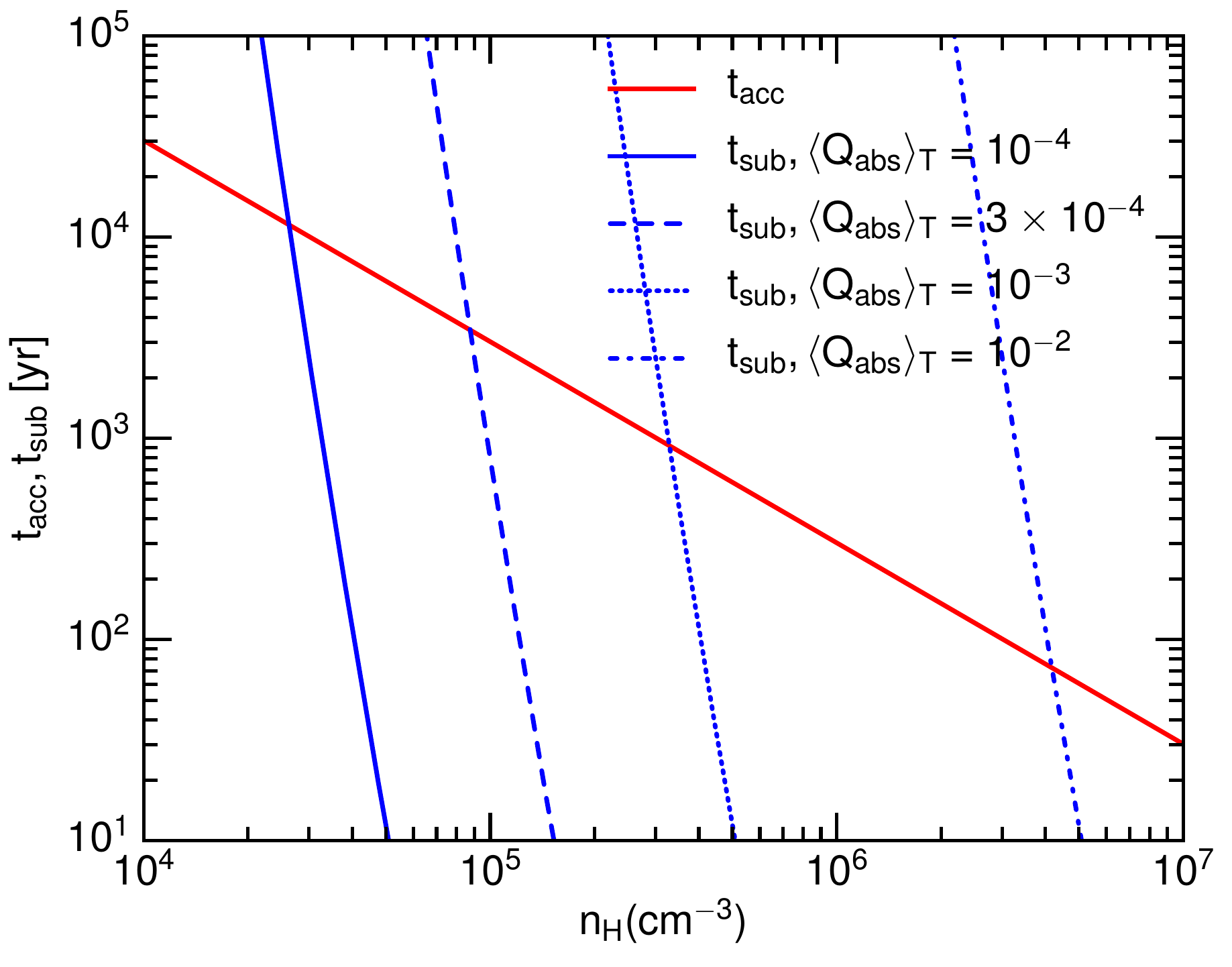}
\caption{Comparison of accretion timescale (red line) with the sublimation time (blue lines) by collisional heating for different emissivities $\langle Q_{\rm abs}\rangle_{T}$, assuming $T_{\gas}=T_{\rm CMB}$ and the grain size of $a=1\mum$. The typical emissivity at low temperatures is $\langle Q_{\rm abs}\rangle_{T}= 10^{-4}$.}
\label{fig:tacc_tsub}
\end{figure}

Figure \ref{fig:tacc_tsub} shows the accretion time and sublimation time as functions of gas density for different emissivities $\langle Q_{\abs}\rangle_{T}$. For the typical value of $\langle Q_{\rm abs}\rangle\sim 10^{-4}$, sublimation is faster than accretion for $n_{\H}>2\times 10^{4}\cm^{-3}$. On the other hand, in lower density regions, accretion is faster than sublimation, but heating by CRs and interstellar radiation could still be important for heating the gas above $T_{\rm CMB}$ and increase sublimation of H$_2$ ice. 

\subsection{Implications: Could `Oumuamua made of H$_2$ ice survive the journey from the birth site to the solar system?}
Assuming that H$_2$ objects could somehow form in the densest regions of GMCs, we found that sublimation by collisional heating inside the GMC would destroy the objects before their escape into the ISM. We also studied various destruction mechanisms of H$_2$ ice in the ISM. In particular, we found that H$_2$ objects are heated by the average interstellar radiation, so that they cannot survive beyond a sublimation time of $t_{\rm sub}\sim 10$ Myr for $R= 300$ m (see Figure \ref{fig:tcom}). Only H$_2$ objects larger than $5\km$ could survive.

\subsection{Implications for H$_2$ ice as baryonic dark matter}
Primordial snowballs were suggested as baryonic dark matter \citep{1996Ap&SS.240...75S}. Previous studies considered collisions between snowballs as a destructive mechanism (\citealt{1986ApJ...303...56H}; \citealt{1999ApJ...516..195C}). \cite{1986ApJ...303...56H} studied destruction of H$_2$ ice by the CMB and found that at redshift $(1+z)=3.5$, sublimation would rapidly destroy H$_2$ ice. Later, \cite{1996Ap&SS.240...75S} argued that the treatment of sublimation by \cite{1986ApJ...303...56H} was inadequate because evaporative cooling was not taken into account. In this work, we have shown that the evaporative cooling is only important for $T_{\rm ice}\gtrsim 3\K$. Even when evaporative cooling is taken into account, thermal sublimation by starlight still plays an important role in the destruction of H$_2$ objects. The present CMB temperature $T_{\rm CMB}$ is not high enough to rapidly sublimate H$_2$ ice. However, at redshifts $z>1$, the CMB temperatures of $T_{\rm CMB}> 5$ K, can rapidly destroy H$_2$ objects of $R\sim 1\km$ within $t_{\rm sub}\sim 48\yr$, based on Equation (\ref{eq:tsub1}).

More importantly, we found that the formation of H$_2$ objects cannot occur in dense GMCs because collisional heating raises the temperature of dust grains, resulting in rapid sublimation of H$_2$ ice mantles. Thus, we find that large objects rich in H$_2$ ice are unlikely to form in dense clouds, in agreement with the conclusions of \cite{1969Natur.224..251G}. 

Lastly, if H$_2$ objects form via a phase transition, as proposed by \cite{2018A&A...613A..64F}, they must be larger than $\sim 5$ km to survive the journey from the GMC to the solar system.

\section{Summary}\label{sec:concl}
We have studied the destruction of H$_2$ ice objects during their journey from their potential birth sites to the solar system. Our main findings are as follows:

\begin{enumerate}

\item Destruction of H$_2$ ice-rich objects by thermal sublimation due to starlight is important, whereas destruction by CRs and interstellar matter is less important.

\item The minimum radius of H$_2$ objects is required to be $R_{\rm min}\sim 5\km$ for survival in the ISM from the nearest GMC.

\item H$_2$ objects of radius $R<200$ m could be destroyed on the way from the GMC to the ISM due to thermal sublimation induced by collisional heating.

\item Formation of H$_2$ ice-rich grains in dense GMCs is unlikely to occur due to rapid sublimation induced by collisional heating. This makes the formation of H$_2$-rich objects improbable.

\end{enumerate}

\acknowledgements
We thank the anonymous referee for a constructive report, as well as, Ed Turner, Ludmilla Kolokolova, Alex Lazarian, and Shu-ichiro Inutsuka for useful comments. T.H. acknowledges the support by the National Research Foundation of Korea (NRF) grants funded by the Korea government (MSIT) through the Mid-career Research Program (2019R1A2C1087045). A.L. was supported in part by a grant from the Breakthrough Prize Foundation.

%--------------adding references-----------------------------------
% or other styles: mcbride,plain, abbrv, acm, alpha, apalike, apj
%\bibliographystyle{/Users/thiemhoang/Dropbox/Papers2/apj}
%\bibliography{/Users/thiemhoang/Dropbox/Papers2/cites_paperApJ,/Users/thiemhoang/Dropbox/Papers2/cites_Books}
\bibliography{ms.bbl}

\end{document}